\documentclass[journal]{IEEEtran}
\usepackage{times} 
\usepackage{mathptmx} 

\ifCLASSINFOpdf
\else
\fi

\usepackage{amsmath,amssymb,amsfonts}
\usepackage[linesnumbered, ruled, vlined]{algorithm2e}
\usepackage{algpseudocode}
\usepackage{graphicx}
\usepackage{textcomp}
\usepackage{lipsum}
\usepackage{amsthm}
\usepackage{siunitx}
\usepackage{amsmath}
\usepackage{subcaption} 
\usepackage{multicol}
\usepackage{multirow}
\usepackage{mathtools}
\usepackage{color}
\usepackage{bm}
\usepackage{indentfirst}
\usepackage{nomencl}
\usepackage{booktabs}
\usepackage{cite}
\usepackage{mathtools}
\usepackage{accents}
\usepackage{tikz}

\usepackage[hyphens]{url}
\usepackage[hidelinks]{hyperref}
\usepackage{makecell}

\SetKwInOut{Input}{Input}
\SetKwInOut{Output}{Output}

\begin{document}

\title{End-to-End Learning-based Operation of Integrated Energy Systems for Buildings and Data Centers}

\author{ 
	Zhenyu~Pu, 
		Yu~Yang,~\IEEEmembership{Member,~IEEE,}
		Liang~Yu,~\IEEEmembership{Senior Member,~IEEE,}
		and~Xiaohong~Guan,~\IEEEmembership{Life Fellow,~IEEE}
		\thanks{This  work  is  supported by the National Natural Science Foundation of China  (62403373, 72595834, 62192752).}
		\thanks{Z. Pu and Y. Yang are with School of Automation Science and Engineering, Xi'an Jiaotong University, Shaanxi, China  (e-mail:  zhenyupu@stu.xjtu.edu.cn, yangyu21@xjtu.edu.cn).
			\textit{Y. Yang is the corresponding author.}
		}
		\thanks{L. Yu is with the College of Automation, Nanjing University of Posts
		and Telecommunications, Nanjing 210023, China (e-mail: liang.yu@njupt.edu.cn).}
		\thanks{X. Guan is with the Center for Intelligent and Networked Systems,
			Department of Automation, Tsinghua University, Beijing 100084, China, and
			also with the Ministry of Education Key Laboratory for Intelligent Networks
			and Network Security Laboratory, Xi'an Jiaotong University, Xi'an 710049,
			China (e-mail: xhguan@tsinghua.edu.cn).}
	}

\markboth{Accepted by Young Adults Congress of China 2026}%
{Shell \MakeLowercase{\textit{et al.}}: Bare Demo of IEEEtran.cls for IEEE Journals}

\maketitle

\begin{abstract}
	Buildings and data centers (DCs) are energy-intensive sectors, playing a critical role to achieve the low-carbon and sustainable energy transition targets. To this end, integrated energy system (IES) that  incorporates   diverse renewables, energy generation, conversion, and storage technologies to enable coordinated multi-energy supply have been widely investigated for both buildings and DCs. However,  few works consider the two sectors jointly within IES to exploit their substantial synergistic benefits.  
	Meanwhile, the operational optimization of IES remains challenging due to the difficulty to  predict the multi-energy demand and supply accurately.
	To address these gaps, this paper investigates IES for coordinated multi-energy supply of  buildings and DC, where the waste heat from DCs is recovered and reused to enhance energy efficiency. Moreover, an end-to-end learning-based method is proposed for the operational optimization of IES under uncertainty. Unlike conventional predict-then-optimize approaches, the proposed method integrates the training of prediction models for  uncertain variables with the  constrained optimization  of IES into a unified learning framework, guiding the training of prediction models to improve  operational performance, rather than prediction accuracy, thereby mitigating the impacts of predictions errors.
	Case studies based on real-world datasets show that the proposed methods improves the operational performance of IES by about 7-9\% compared to existing predict-then-optimize methods. In addition, coordinating buildings and DCs within IES shows substantial economic benefits. In particular,  the  waste heat recovery from DCs leads to approximately 10\%  of  total energy cost reduction of the IES.
\end{abstract}


\IEEEpeerreviewmaketitle

\section{Introduction}  
   
\IEEEPARstart{B}{uildings} remain the dominant energy sector, accounting for around 30–40\% of global energy consumption~\cite{IEA_Buildings}, while data centers (DCs) have emerged as a rapidly growing energy-intensive sector driven by the proliferation of artificial intelligence  and digital services~\cite{DataCentreMag2023efficiency, IEA2025_energy_ai}. Improving their energy efficiency and reducing fossil-fueled energy consumption have been recognized  critical to achieve the global decarbonization and sustainability targets.

To achieve the objective, integrated energy systems (IES), which incorporates  renewables and  diverse generation, conversion, and storage technologies to enable
 coordinated multi-energy supply (i.e., electricity, heating, and cooling), have been widely investigated for both buildings~\cite{comodi2019achieving, cao2018optimal} and DCs~\cite{xu2023optimal, keskin2022optimal}. 
 By enabling intelligent energy management, IES offer significant potential to enhance energy efficiency and facilitate renewable utilization. 
Recently, a growing number of studies have incorporated hydrogen energy storage system into IES  (see~\cite{dong2022optimal, liu2021optimal, fang2022multiple, zhang2023modeling, pu2026representation} for examples), aimed  to leverage the high  long-term and large-scale  storage  efficiency of hydrogen to address the growing supply-demand balancing challenges caused by the increasing penetration of renewables.
Despite these advances, existing studies mainly investigated IES   for buildings and DCs separately, with limited attention to their coordination to exploit the substantial  synergistic benefits. 
\emph{First}, the  complementary energy consumption behaviors of the two  sectors can be explored to  enhance the  operating flexibility. 
\emph{Second}, the substantial waste heat  from DCs   can be recovered  to  satisfy the heating and cooling demands of buildings, thereby improving the overall energy efficiency~\cite{chen2023experimental, hao2025data}. 

While  IES represents a  promising pathway to achieve the energy transition targets,  its operational optimization represents a major challenge.  
It requires to coordinate the operation of diverse energy generation, conversion and storage devices to ensure multi-energy supply-demand balance under uncertainty while achieving minimal energy cost. 
Stochastic programming (SP), reinforcement learning (RL) and model predictive control (MPC) have been widely
 investigated to address this challenge (see \cite{yu2023review, energyManagementOptimizationReview2023} for a comprehensive review).  
 While both SP and RL are computational intensive for practical applications, the prediction-based MPC is often preferred due to relatively low computational cost. 
%
Specifically, MPC generally  computes  the optimal operating strategy of systems based on  the predictions of uncertain variables and only requires to solve deterministic optimization problems. 
%
However, the class of methods often suffers from the deficient operational performance due to the prediction difficulty of uncertain variables. 
Particularly, the multi-energy supply and demand of IES are  influenced  by multiple factors and are hard to  predict accurately. 
%
In recent years, learning-based approaches, such as end-to-end predictive control~\cite{donti2017task}, decision-focused prediction~\cite{shah2022decision, mandi2024decision}, and predict-to-optimize~\cite{SPO}, have emerged as promising framework to address the effects of prediction errors. 
These methods integrate the predictions of uncertain variables with downstream constrained optimization in a unified framework and directly target on the operational performance, rather than the prediction accuracy.
They are motivated by the awareness that 
prediction accuracy does not necessarily translate into enhanced operational performance, despite their close correlation.
%
While learning-based approaches have been demonstrated promising, their practical implementation remains to be investigated. The challenges  mainly arise from the  non-analytical nature of most constrained optimization, making it difficult to back propagate gradients from decision performance to prediction model parameters.
Recently, advances in differentiable optimization offer new opportunities to address these challenges. A recent work on implicitly defined layers enables differentiable constrained optimization as a neural layer based on the Karush-Kuhn-Tucker (KKT) conditions and the implicit function theorem~\cite{implicitly}.
In addition, differentiable convex optimization layers (i.e., CVXPYlayers) support automatic differentiation through convex solvers~\cite{cvxpylayer_proposed, cvxpylayer_applied}. These advances make it possible to integrate predicting models with constrained optimization in an end-to-end learning framework.

\textbf{Motivated by the above researches, this paper makes contributions from two aspects.} \emph{First}, we investigate a hydrogen-based IES for coordinated multi-energy supply (i.e., electricity, heating and cooling) of 
	buildings and DCs, where the waste heat from DCs is recovered and reused to meet the heating and cooling demands of the two sectors. 
	\emph{Second}, we propose an end-to-end learning-based  approach for the operational optimization of IES  under uncertainty. The method integrates the training of prediction model for uncertain variables and constrained optimization for 
	IES operation in a unified framework,  guiding the training of prediction models to improve the operational performance rather than prediction accuracy, thereby mitigating the effects of prediction errors. 
Case studies based on real-world datasets show that  the proposed end-to-end learning-based approach can improve the performance by about 7-9\% over conventional predict-then-optimization scheme (i.e., decoupled). This performance improvements are noteworthy as they are achieved only through changing the learning pipeline instead of modifying the prediction models.  
Besides,  the coordination of buildings and DCs within IES show substantial economic benefits. Particularly, the  recovered heat from DCs can be effectively utilized to satisfy heating and cooling demands, leading to approximately 10\% total energy cost reduction of the IES.

The rest of the paper is  as follows: Section~\ref{sec:problem formulation} introduces the IES and its mathematical formulation for operational optimization. Section~\ref{sec:Methodology}  proposes the end-to-end learning-based method. Section~\ref{sec:Experiments and Results} evaluate the operational performance of the system. Section~\ref{sec:conclusion} concludes the work.

\begin{figure}[htbp]
    \centering
    \includegraphics[width=0.9\linewidth]{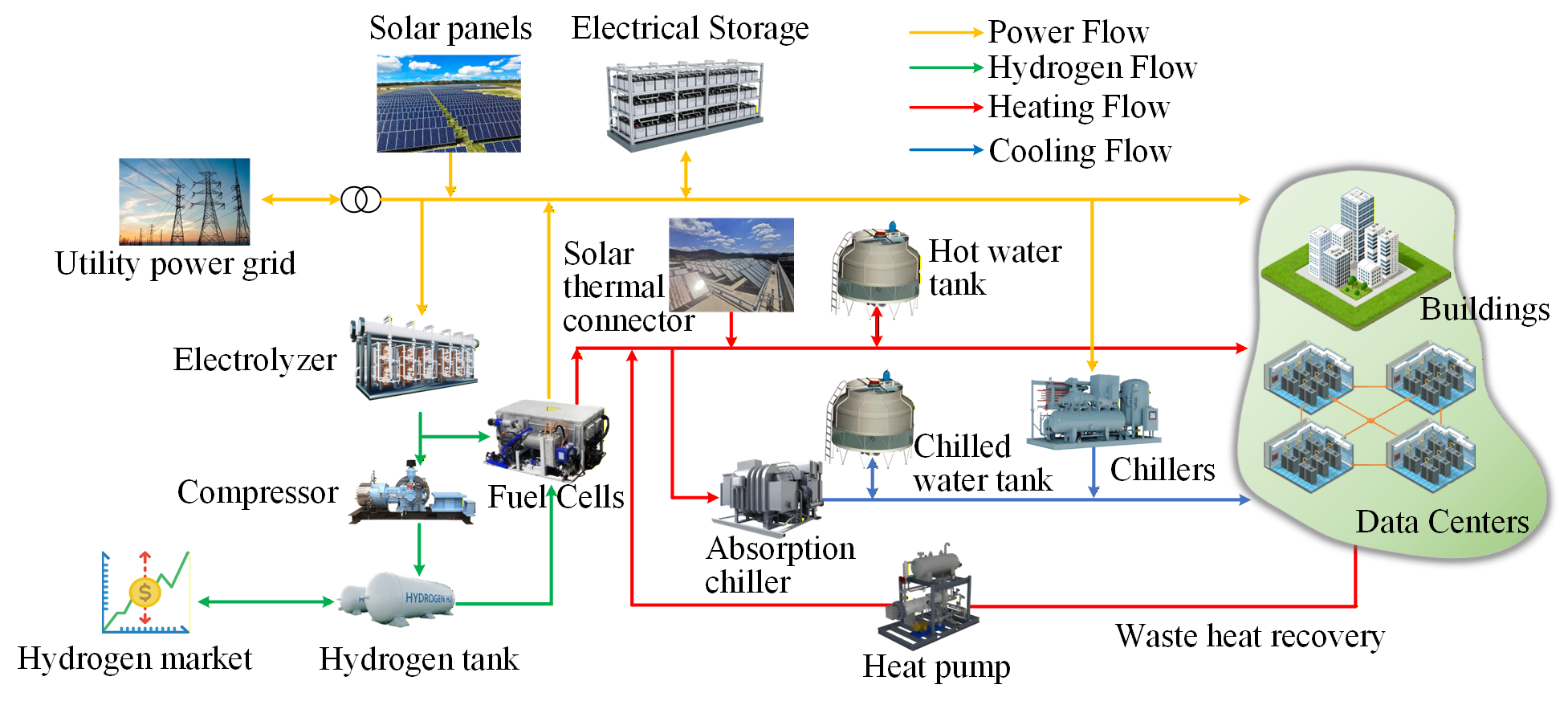}
    \captionsetup{belowskip=-20pt}
	\caption{hydrogen-based IES for buildings and DCs}
    \label{fig:system}
\end{figure}

\section{The Problem}
\label{sec:problem formulation}

\subsection{Hydrogen-based IES for buildings  and data centers (DCs)}
 As illustrated in Fig.~\ref{fig:system}, a hydrogen-based IES for  supplying electricity, heating, and cooling energy to buildings and DCs is considered. 
Buildings primarily consume electricity, domestic hot water, heating or cooling energy for space conditioning, and DCs mainly consume electricity and cooling energy to power and cool  IT equipment.
 The IES is connected to utility power grid and hydrogen market for primary energy resources.  Solar panels and  thermal collectors are deployed for harvesting on-site renewable energy. 
Diverse energy generation, conversion and storage devices are integrated in the IES. 
The storage devices include  an electrical battery storage,  a hot and chilled water tank, and a hydrogen energy storage system (HESS). The HESS consists of electrolyzers (ELs), compressor, hydrogen tank (HT), and fuel cells (FCs).   
 Hydrogen can be produced by ELs via renewable-powered water electrolysis or purchased from 
 hydrogen market. Hydrogen can be 
 stored in the tank and  converted into electricity and heat by FCs when needed.  
The dissipated heat of FCs is  recovered by a heat recovery unit and used to produce heat water or   converted into cooling energy by absorption chiller (AC). 
To improve energy efficiency,   the waste heat of DCs is also recovered from 
DC cooling systems.  Considering the waste heat temperature generally ranges
25-50~$^\circ\mathrm{C}$ and 40-80~$^\circ\mathrm{C}$ 
for air-cooled and liquid-cooled systems, representing low-grade heat~\cite{hao2025data}, heap pump (HP)  is  deployed to upgrade the heat, enabling its effective utilization  by the AC. 

\vspace{-12pt}

\subsection{Problem Formulation}

We consider the operational optimization of the IES for  minimizing energy purchase cost. 
The problem is considered in a discrete-time framework with the scheduling horizon $T$ equally divided into decision intervals of $\Delta t$. 
The mathematical formulation  for IES operation based on predicted multi-energy supply and demand is presented as follows.

\textbf{Electrical storage system (ESS)}:  The ESS corresponds to  lithium batteries, whose operating dynamics and physical limits over the scheduling horizon  are modeled as 
\begin{subequations} \label{eq:ESS}
	\begin{align}
		&~S_{t + \Delta t}^{\rm ess} = S_{t}^{\rm ess}
		+ \big(P_{t}^{\rm ess, ch} \eta^{\rm ess, ch}
		-{P_{t}^{\rm ess,  dis}}/{\eta^{\rm ess,  dis}}\big)\Delta t, 
		\label{eq:ess_energy_dynamics} \\
		&~~0 \le P_{t}^{\rm  ess, ch}, P_{t}^{\rm ess,  dis} \le P^{\rm ess}_{\max}, \label{con:ess_charge_limits}\\
		&~~S^{\rm ess}_{\min} \le S_{t}^{\rm ess} \le S^{\rm ess}_{\max}, \quad t \in T. 		\label{con:ess_energy_range} 
	\end{align}
\end{subequations}
where $t$ denotes the time.  $S_{t}^{\rm ess}$ [kWh] denotes the stored electricity. $P_{t}^{\rm ess,\rm ch}, P_{t}^{\rm ess,\rm dis}$ [kW]  are scheduled charging and discharging power, with  $\eta^{\rm ess,ch}, \eta^{\rm ces,  dis}$ denoting charging and discharging efficiencies. Eq.~\eqref{eq:ess_energy_dynamics} models the charging dynamics of ESS.  
Cons.~\eqref{con:ess_charge_limits} model the charging and discharging rate limits.  Cons.~\eqref{con:ess_energy_range} define the permissible operating range of energy capacity.  It is worthy noting that the non-simultaneous charging and discharging constraints for  all storage devices are omitted without affecting the solution due to the much higher purchasing price than selling price of utility grid.

\textbf{Hot water tank}:  The hot water tank is used  as thermal energy storage (TES) for satisfying  diverse forms of heating demand. 
The stored energy dynamics  and operating limits  are 
captured by 
\begin{subequations}
	\begin{align}
		&S_{t+\Delta t}^{\rm tes} = S_{t}^{\rm tes}
		+ \Big(g_{t}^{\rm tes, ch} \eta^{\rm tes, ch}
		-{g_{t}^{\rm tes,  dis}}/{\eta^{\rm tes,  dis}}\Big)\Delta t, 
		\label{eq:tes_energy_dynamics} \\
		&~~0 \le g_{t}^{\rm  tes, ch}, g_{t}^{\rm tes,  dis} \le g^{\rm tes}_{\max},
		\label{con:tes_charge_limits} \\
	    &~~S^{\rm tes}_{\min} \le S_{t}^{\rm tes} \le S^{\rm tes}_{\max}, \quad  t \in T.
		\label{con:tes_energy_range} 
	\end{align}
\end{subequations}
where $S_{t}^{\rm tes}$ [kWh] represents the stored heating energy in hot water tank. $g_{t}^{\rm tes, ch}, g_{t}^{\rm tes, dis}$ [kW] are scheduled charging and discharging heating energy, with  $\eta^{\rm tes, ch}$ and $\eta^{\rm tes, dis}$ denoting the corresponding efficiencies. Eq.~\eqref{eq:tes_energy_dynamics} models the dynamics of stored heating energy in the tank.
Cons.~\eqref{con:tes_charge_limits} model  charging and discharging rate limits. 
 Cons.~\eqref{con:tes_energy_range} defines the permissible energy capacity operating range of hot water tank

\textbf{Chilled Water Tank}:  The chilled water tank is used  as cooling energy storage (CES) for satisfying  diverse forms of cooling demand. The stored energy dynamics  and operating limits  are captured by 
\begin{subequations}
	\begin{align}
		&S_{t + \Delta t}^{\rm ces} = S_{t}^{\rm ces}
		+ \Big(q_t^{\rm ces, ch} \eta^{\rm ces, ch}
		- {q_t^{\rm ces,  dis}}/{\eta^{\rm tes,  dis}}\Big)\Delta t, 
		\label{eq:ces_energy_dynamics} \\
		&~~0 \le q_{t}^{\rm  ces, ch}, q_{t}^{\rm ces,  dis} \le q^{\rm ces}_{\max},
		\label{con:ces_charge_limits} \\
		&~~S^{\rm ces}_{\min} \le S_{t}^{\rm ces} \le S^{\rm ces}_{\max}, \quad  t \in T.
		\label{con:ces_energy_range}
	\end{align}
\end{subequations}
where $S_{t}^{\rm ces}$ [kWh] denotes the stored cooling energy in the tank. $q_{t}^{\rm ces,  ch}, q_{t}^{\rm ces,  dis}$ [kW] are  scheduled charging  and discharging cooling energy, with $\eta^{\rm ces,  ch}, \eta^{\rm ces,  dis}$ denoting the  efficiencies. Eqs.~\eqref{eq:ces_energy_dynamics} model the  dynamics of stored cooling energy in the tank.  Cons.~\eqref{con:ces_charge_limits} model the charging and discharging rate limits. Cons.~\eqref{con:ces_energy_range} specify the permissible energy capacity  operating range of chilled water tank.  

\textbf{Hydrogen energy storage system (HESS)}:
HESS consists of ELs, a compressor, a hydrogen tank, and FCs. The hydrogen production of ELs depends on the injected electrical power and can be approximated by a linear model~\cite{liu2021optimal}
\begin{equation}
	\begin{split}
		& m_{t}^{\rm EL} = k^{\rm EL} \cdot P_{t}^{\rm EL} \cdot \eta, \\
		& P_{t}^{\rm EL} \leq P^{\rm EL,\max}, \quad \forall t \in T.
	\end{split}
\end{equation}
where $m_{t}^{\rm EL}$ [kg/h] is the hydrogen production rate of ELs at  time $t$, $P_{t}^{\rm EL}$ [kW] is the input electrical power of ELs, and $k^{\rm EL}$ [kg/kWh] denotes the electricity-to-hydrogen conversion coefficient.
$\eta$ represents the efficiency of the compressor used for hydrogen purification and compression, and $P^{\rm EL,\max}$ denotes the maximum allowable electrical input power of ELs.
FCs  convert hydrogen into electricity and heat, whose operating characteristics can be modeled as 
\begin{equation}
\begin{split}
& P_{t}^{\rm FC} = k^{\rm FC} \cdot m_{t}^{\rm FC}, \\
& g_{t}^{\rm FC} =  \eta^{\rm rec} \cdot \theta \cdot P_{t}^{\rm FC}, \quad \forall t \in T. 
\end{split}
\end{equation}
where $P_{t}^{\rm FC}$ and $g_{t}^{\rm FC}$ denote the produced electricity and heat by the FCs. $\theta$ denote the electricity to heat ratio of FCs. $\eta^{\rm rec}$ denotes the efficiency of heat recovery unit equipped with FCs. 
Hydrogen tank is used to store hydrogen,  its operating dynamics and constraints can be described as 
\begin{equation}
\begin{split}
	& m_{t +\Delta t}^{\rm tank} = m_{t}^{\rm tank} + (m_{t}^{\rm EL} - m_{t}^{\rm FC} + m_{t}^{\rm buy}) \cdot \Delta t, \\
	&0 \leq  m_{t}^{\rm tank} \leq m^{\rm tank, \max} \\
	& 0 \leq m_{t}^{\rm buy} \leq m^{\rm buy, \max},  \quad \forall t \in T.
\end{split}
\end{equation}
where $m_t^{\rm tank}$ [kg] denotes the stored hydrogen at time $t$. $m_t^{\rm buy}$ [kg/h] denotes the purchased  hydrogen from hydrogen market, which is non-negative and upper bounded by $m_t^{\rm buy, \max}$. $m^{\rm tank, \max}$ denotes the capacity of hydrogen tank.

\textbf{Absorption Chiller (AC)}:  AC converts heat into cooling energy. The produced cooling power  is determined by the injected heat  and energy conversion efficiency of AC, which can be modeled as  
\begin{equation}
	\begin{split}
		&q^{\rm ac}_{t} = g^{\rm ac}_{t} \cdot \eta^{\rm ac},  \\
		& g^{\rm ac}_{t} \leq g^{\rm ac}_{\max}, \quad t \in T. 
	\end{split}
\end{equation}
where $g^{\rm ac}_{t}$ [kWh] denotes the injected heat to AC at  time  $t$. 
$q_{t}^{\rm ac}$ [kWh] is produced cooling energy. $g^{\rm ac}_{\max}$ captures the operating limits of the AC.

\textbf{Chillers}:  Considering the possible intensive cooling demand, electrical chillers are deployed within the IES to produce chilled water. Their operating characteristics can be  modeled as 
\begin{equation}
\begin{split}
	& q_{t}^{\rm chiller} = {\rm COP} \cdot P_t^{\rm chiller}, \\
	&  0 \le P_{t}^{\rm chiller} \le P^{\rm chiller, \max}, \quad \forall t \in T. 
\end{split}
\end{equation}
where $q_t^{\rm chiller}$ [kW] denotes the produced cooling energy by the chillers. $P_t^{\rm chiller}$ [kW] denote the injected electrical power to chillers. COP captures the coefficient of performance (COP) of chillers. $P^{\rm chiller, \max}$ denotes the maximum  electrical power injection of the chillers. 

\textbf{Solar panels}: Solar panels convert solar radiation into electrical power, and the following model is commonly used to characterize its predicted output
\begin{equation}
	P_{t}^{\rm solar} = \eta^{\rm pv} \cdot A_{\rm pv} \cdot \hat{Q}_{t}^{\rm rad}, \quad \forall t \in T. 
\end{equation}
where $P_{t}^{\rm solar}$ [kW] is generated solar power, determined by solar panel conversion efficiency $\eta^{\rm pv}$, solar collection area $A_{\rm pv}$ [m$^2$], and the predicted solar radiation  $\hat{Q}_{t}^{\rm rad}$ [kW/m$^2$].

\textbf{Solar Thermal Connector}:  Solar thermal collector converts solar radiation it into heat, which can be modeled as
\begin{equation}
	g_{t}^{\rm solar} = \eta^{\rm stc} \cdot A_{\rm stc} \cdot \hat{Q}_{t}^{\rm rad}, \quad \forall t \in T. 
\end{equation}
where $g_{t}^{\rm solar}$ [kW] is absorbed heat by solar thermal collector at time  $t$, which is determined by its energy conversion efficiency $\eta^{\rm stc}$,  collector area $A_{\rm stc}$  [$\rm m^2$] and the predicted solar radiation $\hat{Q}_{t}^{\rm rad}$  [kW/m$^2$]. 

\textbf{Heat pump (HP)}: HP converts the low-grade waste heat of data centers into high-grade heat, which can be modeled as 
\begin{equation}
	g^{\rm HP}_{t} =\eta^{\rm HP} \hat{g}_{t}^{\rm DC}, \quad \forall t \in T. 
\end{equation}
where $\hat{g}_{t}^{\rm DC}$ is  predicted waste heat from the DCs at   time $t$.  $\eta^{\rm HP}$ denotes conversion efficiency of HP.

\textbf{Energy Balances}:
The fundamental function of the IES is to ensure the  balance of multi-energy supply and demand across the time, which can be captured by the following equations 
\begin{subequations} \label{eq:energy_balance}
	\begin{align}
		&P^{\rm g,  buy}_{t} - P^{\rm g, sell}_{t} + P_{t}^{\rm solar} +  P_{t}^{\rm FC}= P_t^{\rm ess,ch} + P_{t}^{\rm ess,dis}  \label{eq:power_balance}   \\
		& \quad \quad \quad \quad \quad \quad \quad \quad \quad   +  P_{t}^{\rm EL} + P_t^{\rm chiller} + \hat{P}_{t}^{\rm B} + \hat{P}_{t}^{\rm DC}, \notag \\
		&g_{t}^{\rm solar} + g_{t}^{\rm FC} +g_{t}^{\rm HP} = g_{t}^{\rm tes,ch} + g_{t}^{\rm tes,dis}+ g_t^{\rm ac}+ \hat{g}_{t}^{\rm B},   \label{eq:heat_balance}\\
		& q_{t}^{\rm ac} + q_{t}^{\rm chiller}= q_{t}^{\rm ces,ch} + q_{t}^{\rm ces,dis} + \hat{q}_{t}^{\rm B} + \hat{q}_{t}^{\rm DC}, \forall t\in T. \label{eq:cool_balance}
	\end{align}
\end{subequations}
where   $\hat{P}_{t}^{\rm B}, \hat{g}_{t}^{\rm B}$, $\hat{q}_{t}^{\rm B}$ are predicted  electricity, heating, and cooling  demand of all buildings.  $P^{\rm g, buy}_{t}, P^{\rm g, sell}_{t}$ [kW] denote the electricity purchased from and sold to the utility grid at  time  $t$, which are non-negative and upper bounded
\begin{equation} \label{eq:grid_trading}
\begin{split}
& 0 \le P^{\rm g,  buy}_{t}, P^{\rm g,  sell}_{t}\le P^{\rm g,  \max}, \forall t \in T.  
\end{split}
\end{equation}

\textbf{Objective}: The operational optimization objective is to minimize the operating cost  over the scheduling horizon $T$, which  can be modeled as 
\begin{equation}
	L^{\rm cost} \!\!=\! \sum_{ t \in T} \left[\lambda^{\rm b}_{t} P_t^{\rm g, buy} \!-\! \lambda^{\rm s}_{t} P_t^{\rm g, sell}  + \lambda^{\rm h}_{t} m^{\rm buy}_{t}  \right]\Delta t
	\label{eq:objective}
\end{equation}
where $\lambda^{\rm b}_{t}$ and $\lambda^{\rm s}_{t}$ are  buying and selling price of utility grid at  time  $t$. $\lambda^{\rm h}_{t}$ is  hydrogen trading price (buying and selling). 
The first two terms are electricity transaction cost, and the third is hydrogen transaction cost.

\textbf{Constrained optimization}: The operational optimization of the IES over the scheduling horizon $T$   corresponds to  the following constrained optimization 
\begin{equation} \label{eq:p}
	\begin{split}
	\min 	&~ L^{\rm cost} \!\!=\! \sum_{ t \in T} \left[\lambda^{\rm b}_{t} P_t^{\rm g, buy} \!\!-\!\! \lambda^{\rm s}_{t} P_t^{\rm g, sell}  + \lambda^{\rm h}_{t} m^{\rm buy}_{t}  \right]\Delta t \\
		& {\rm s.t.}~\eqref{eq:ESS}-\eqref{eq:grid_trading}.
	\end{split}
	\tag{$\mathbf{P}$}
\end{equation}
The problem depends on the predictions of   multi-energy supply and demand: $[\hat{Q}_t], [\hat{P}^{\rm B}_t], [\hat{g}^{\rm B}_t]$, $[\hat{q}^{\rm B}_t]$, $[\hat{P}^{\rm DC}_t],  [\hat{g}^{\rm DC}_t], \forall t \in T$. However, these uncertain variables are intrinsically influenced by multiple factors, making them difficult to  predict accurately. This often  leads to the deficient operational performance with most existing prediction-based  approaches. 

\vspace{-8pt}

\section{End-to-end Learning-based Method}
\label{sec:Methodology}
To address the challenges, this section  proposes an end-to-end learning-based approach for the operational optimization of IES under uncertainty.
Different from conventional predict-then-optimize scheme, this method integrates the training of predictive models for uncertain variables with the constrained optimization \eqref{eq:p} for IES operation in an end-to-end learning framework,  guiding the  training of prediction models to improve operational performance rather than prediction accuracy. The framework consists of two modules: predictive models for uncertain variables and the constrained optimization for decision-making as introduced below. 

\begin{figure}[htbp]
	\centering
	\includegraphics[width=0.9\linewidth]{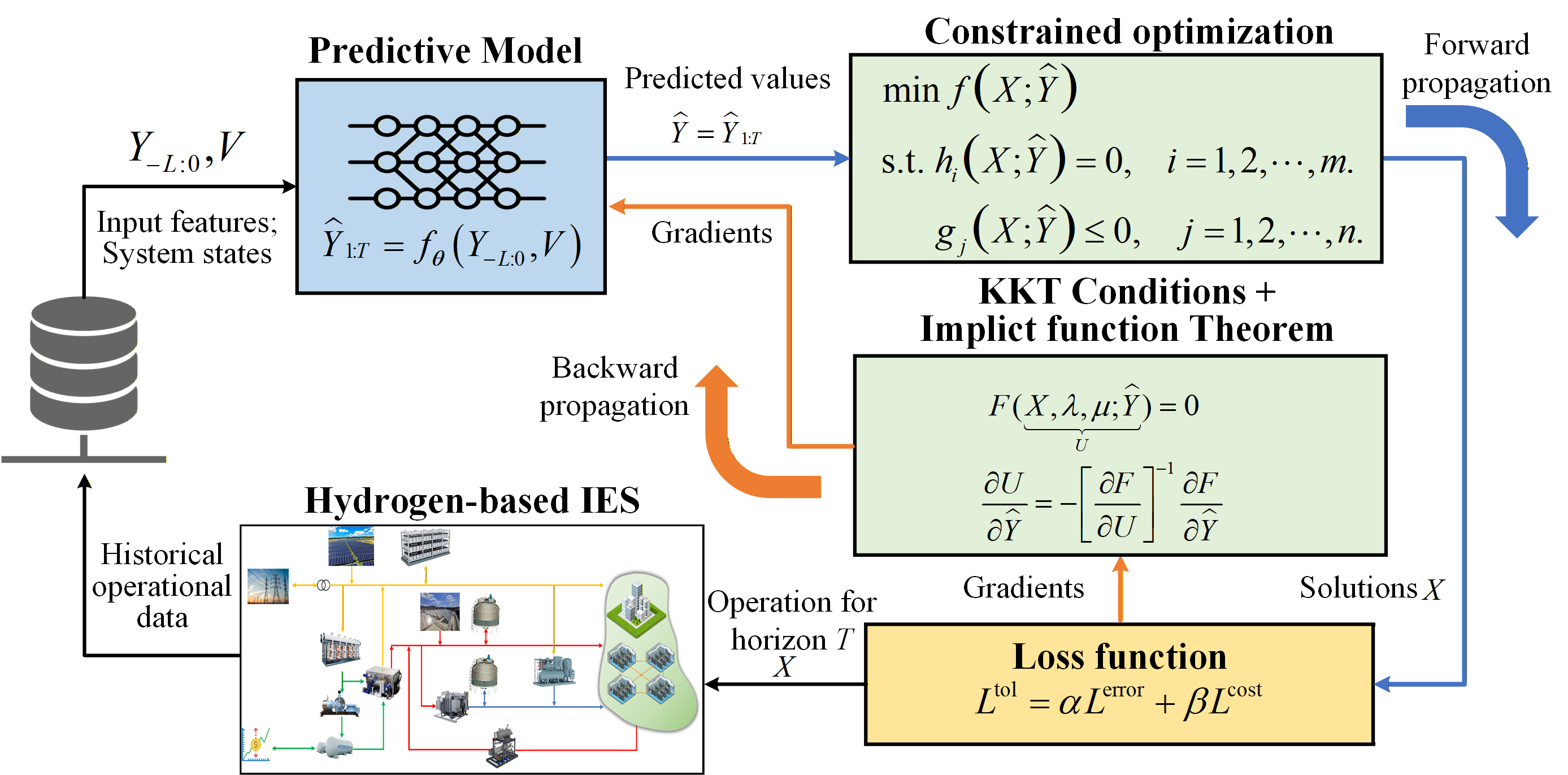}
	\captionsetup{belowskip=-20pt}
	\caption{End-to-end learning-based method for IES}
	\label{fig:method_architecture}
\end{figure}

\vspace{-3mm}
\subsection{Prediction Model}
We denote the predictions of uncertain variables  over the scheduling horizon $T$ as 
\begin{equation}
\begin{split}
\hat{\bm{Y}}_{1:T}  = \left\{ \hat{Q}_{1:T}^{\rm rad}, \hat{P}_{1:T}^{\rm B}, \hat{g}_{1:T}^{\rm BC}, \hat{q}_{1:T}^{\rm B}, \hat{P}_{1:T}^{\rm DC}, \hat{g}_{1:T}^{\rm DC}\right\}. 
\end{split}
\end{equation}
where $1\!:\!T$ denotes time indices of the scheduling horizon $T$.

To enable prediction-based optimization, the uncertain variables  over the scheduling horizon $T$ requires to be forecast.  This  corresponds to the following multi-variable and multi-step forecasting task
\begin{equation} \label{eq:multi-step-forecast}
	\begin{aligned}
		\hat{\bm{Y}}_{1:T}= f_{\bm{\theta}}( \bm{Y}_{-L:0},  \bm{V})
	\end{aligned}
\end{equation}
where $ \bm{Y}_{-L:0}$ denotes past $L$ historical observations up to the present time $0$. $\bm{V}$ denotes extra features to enhance prediction accuracy, such as time-of-day, day-of-week or weather conditions.  $f_{\bm{\theta}}(\cdot)$ presents the forecasting model with $\bm{\theta}$ denoting model parameters. 
Any multi-variable and multi-step forecasting methods are applicable to the forecasting task \eqref{eq:multi-step-forecast}.

\vspace{-8pt}

\subsection{Constrained Optimization as a Neural Layer}

To enable end-to-end prediction and optimization, we develop a framework to  incorporate the constrained optimization \eqref{eq:p}  as a neural layer. 
To achieve the objective, we first restate  Problem \eqref{eq:p} in the compact format:
\begin{equation}\label{eq:P1}
\begin{split}
 \min&~~ f\big(\bm{X};~ \hat{\bm{Y}}\big) \\
 {\rm s.t.}~&h_{i}\big(\bm{X};~ \hat{\bm{Y}}\big) = 0, \quad i = 1, 2, \cdots, m.\\
& g_{j}\big(\bm{X}; ~\hat{\bm{Y}}\big) \le 0, \quad j = 1, 2, \cdots, n. \\
\end{split}
\tag{$\mathbf{P}^\prime$}
\end{equation}
where $\bm{X}:= \bm{X}_{1:T}$ denotes the operational policy, and $\hat{\bm{Y}}: = \hat{\bm{Y}}_{1:T}$ are predictions of the uncertain variables  over the scheduling  horizon $T$. $f(\cdot)$, $h_{i}(\cdot)$, and $g_{j}(\cdot)$ are objective function, equality constraints and inequality constraints of Problem \eqref{eq:p}. 

Problem \eqref{eq:P1} is a convex and differential optimization problem, which satisfies strong  duality.
Therefore, its optimal solutions satisfy the following necessary Karush-Kuhn-Tucker (KKT) equations
\begin{equation}
\begin{cases}
	\nabla_{\bm{X}} f(\bm{X}; \hat{\bm{Y}})
	\!+\! \sum_{i=1}^m \lambda_{i} \nabla_{\bm{X}} h_{i}(\bm{X};\hat{\bm{Y}})\!+\! \sum_{j=1}^n \mu_{j}  \nabla_{\bm{X}} g_{j}(\bm{X};\hat{\bm{Y}}) = \mathbf{0},  \\
	h_{i}(\bm{X}; \hat{\bm{Y}}) = 0, \quad \quad i=1,\dots,m \\
	\mu_{j}\, g_j(\bm{X}; \hat{\bm{Y}}) = 0, \quad j=1,\dots,n
\end{cases}
\label{eq:KKT}
\end{equation}
where $\bm{\lambda}:=\lambda_{1:n}$ and $\bm{\mu}:=\mu_{1:m}$ are Lagrangian multipliers. 
Note that in \eqref{eq:KKT}, the non-negativity condition $\mu_{j} \ge 0$ and the primal feasibility $g_j(\bm{X};~\hat{\bm{Y}}) \le 0$  are omitted,  and the above are necessary but not sufficient  KKT conditions. This is to facilitate the subsequent application of an implicit function theorem. 
Specifically,  \eqref{eq:KKT} corresponds to an implicit function
\begin{equation}
\begin{split}
F(\bm{X}, \bm{\lambda}, \bm{\mu}; \hat{\bm{Y}}) = 0
\end{split}
\end{equation}

We define $\bm{U}:= [\bm{X}, \bm{\lambda}, \bm{\mu}]$ as the primal and dual solutions of problem \eqref{eq:P1}.  According to the implicit function theorem stated in \cite{implicitly}, the derivatives of  solution $\bm{U}$ to the input predictions $\hat{\bm{Y}}$ can be calculated by
\begin{equation} \label{eq:gradient}
	\frac{\partial \bm{U}}{\partial \hat{\bm{Y}}} = - \left[ \frac{\partial F}{\partial \bm{U}} \right]^{-1} \frac{\partial F}{\partial \hat{\bm{Y}}}
\end{equation}


To achieve the objective,  we define the loss function of the learning-based predictive control  as the weighted   forecasting error and operational cost over the scheduling horizon $T$
\begin{equation} \label{eq:training_loss}
	L^{\rm loss}= \alpha L^{\rm error}  + \beta L^{\rm cost}
\end{equation}
where $\alpha, \beta \in [0,1]$ and $\alpha + \beta  = 1$ are weighting coefficients that balance the two components.  The prediction error  is important to stabilize the training of predicting model, which is calculated as $L^{\rm error} = \sum_{t \in T} \Vert \hat{\bm{Y}}_t -\bm{Y}_t \Vert^2$, with
$\bm{Y}_t$ denoting the actual value of the uncertain variables at time $t$. 
The operational cost is involved to guide the training of the predicting model for supporting the operational optimization of the IES system. 

To enable  end-to-end training, the gradients of the loss function need to be back-propagated to the parameters of the prediction model. This can be evaluated by the chain rules
\begin{equation} \label{eq:chain_rule}
\frac{\partial L^{\rm loss}}{\partial \theta} = 
\frac{\partial L^{\rm loss}}{\partial \bm{U}} \cdot  \frac{\partial \bm{U}}{\partial \hat{\bm{Y}}} \cdot \frac{\partial \hat{\bm{Y}}}{\partial \theta}    
\end{equation}
where ${\partial L^{\rm loss}}/{\partial \bm{U}_t}$ and ${\partial \hat{\bm{Y}}}/{\partial \theta}$ can be obtained through Automatic Differentiation. 
The term ${\partial \bm{U}}/{\partial \hat{\bm{Y}}}$ can be obtained by  \eqref{eq:gradient}. Particularly, if Problem \eqref{eq:p} is quadratic programming (QP), 
the CVXPYLayers \cite{cvxpylayer_proposed} developed by Stephen Boyd Research Group can be directly  for back-propagation.
 
The implementation of  end-to-end learning-based method for IES is  illustrated in Fig.~\ref{fig:method_architecture}, which is composed of the forward and backward pipeline. 
 For \textbf{forward propagation}, the predictive model generates forecasts $\hat{\bm{Y}}_{1:T}$ for the scheduling horizon $T$, which are fed into a constrained optimization problem to obtain the optimal decisions $\bm{X}$. The resulting strategy is evaluated by the loss function in \eqref{eq:training_loss}. For \textbf{backward propagation}, gradients of the loss are backpropagated to update the predictive model parameters via \eqref{eq:gradient} and \eqref{eq:chain_rule}. After training, the learned model, together with an optimization solver, can be deployed for  IES operation.

\vspace{-8pt}

\section{Case Studies}
\label{sec:Experiments and Results}
This section evaluates the end-to-end learning-based approach for  IES operation under uncertainty. 
Besides, we investigate the economic benefits of coordinate buildings and DCs in the IES, in particularly how the waste heat recovery from the DCs affects the total energy cost of the system.

\subsection{Simulation setup}
We use real-world datasets to set up case studies.  Electricity prices, multi-energy demands (i.e., electricity, heating, and cooling) of buildings and solar radiation are adopted from CityLearn~\cite{citylearn_v2}. Electricity and cooling demand, as well as  waste heat generation of DCs are taken from the Hewlett Packard Enterprise-Cray EX Frontier data center~\cite{sun2024energy}. 
These raw data are preprocessed by removing missing values and then resampled at 1h resolution.
The scheduling horizon is set to one day and divided into 1h decision intervals (i.e., $T = 24$).
Hydrogen market is omitted in this paper without loss of generality.
For forecasting tasks, 24-hour historical data  together with the time  (i.e., month, day type, hour) are used as input features to predict the uncertain variables for the following day. The dataset spans two months and is split into training, validation, and testing sets by a ratio of 7:1:2.
For DCs, only electricity demand and waste heat are predicted,  and cooling demand is directly obtained by  the fixed power usage effectiveness (PUE).
The widely-used long short-term memory (LSTM) neural network~\cite{LSTM} is adopted as the forecasting model (layers: 2, hidden dim: 128). A joint prediction of multiple uncertain variables is performed to exploit their correlations to enhance prediction accuracy.
For the end-to-end learning-based approach, the batch size is set as 64, and the Adam optimizer is used for model trading.
Both prediction and optimization are implemented using the CVXPY framework, with CVXPYLayer~\cite{cvxpylayer_proposed} integrated into PyTorch for automatic differentiation.
The configurations of energy generation, conversion, and storage devices of IES follow\cite{dong2022optimal, liu2021optimal}.  For the proposed end-to-end approach, dynamic weighting factors are adopted to enhance the learning performance. Specifically,   $\alpha$ decreases linearly from 1.0 to 0.6, and $\beta$ increases from 0 to 0.4 during the training process.

\begin{table}[htbp]
	\centering
	\caption{Prediction and operation performance of IES under different methods.}
	\setlength{\tabcolsep}{2pt}
	\label{tab:performance}
	\begin{tabular}{llcccccc}
		\hline
		\multirow{3}{*}{\centering Cases} & \multirow{3}{*}{\centering Methods} & \multicolumn{4}{c}{Prediction} & \multicolumn{2}{c}{Operation} \\
		\cmidrule(lr){3-6} \cmidrule(lr){7-8}
		& & MAPE & RMSE  & $R^2$ & \makecell[c]{Regression\\ loss [kW]} & \makecell[c]{Total \\cost [\$]} & \makecell[c]{Redu.\\~[\%]} \\
		\hline
		\multirow{3}{*}{Case 1} & \textbf{Optimal}        & --    & --     & -- & -- & 0 & --   \\
		& \textbf{Decoupled}   & 0.307 & 0.089  & 0.916 & 3.40    & 54.76 & --   \\
		& \textbf{End-to-End}  & 0.306 & 0.088  & 0.916 & 3.25    & 49.94 & 8.8 \\
		\hline
		\multirow{3}{*}{Case 2} & \textbf{Optimal}     & --    & --       & --      & -- & 1510.00 & --   \\
		& \textbf{Decoupled}   & 0.307 & 0.089 & 0.916 & 16.98   & 2191.67 & --   \\
		& \textbf{End-to-End}  & 0.328 & 0.101 & 0.891 & 18.72   & 2015.16 & 8.1 \\
		\hline
		\multirow{3}{*}{Case 3} & \textbf{Optimal}     & --    & --    & --         & -- & 4686.74 & --   \\
		& \textbf{Decoupled}   & 0.307 & 0.089  & 0.916 & 33.95   & 6184.75 & --   \\
		& \textbf{End-to-End}  & 0.315 & 0.090 & 0.912 & 34.01   & 5732.46 & 7.3 \\
		\hline
		\multirow{4}{*}{Case 4} & \textbf{Optimal}     & --    & --      & --      & --& 9023.28 & --   \\
		& \textbf{Decoupled}   & 0.307 & 0.089 & 0.916 & 50.93   & 11604.70 & --   \\
		& \textbf{End-to-End}  & 0.311 & 0.089 & 0.916 & 50.34   & 10784.48 & 7.1 \\
		\hline
	\end{tabular}
	\vspace{-16pt}

\end{table}

\subsection{Performance of end-to-end learning-based approach}
We first investigate the performance of the proposed end-to-end  approach for  IES operation. We compare the following methods in case studies. 

\begin{itemize}
	
	\item \textbf{Optimal}: The optimal operating strategy of the IES over the scheduling horizon $T$ is obtained by solving the constrained optimization \eqref{eq:p} based on the actual  renewable supply and multi-energy demand, which serves as the theoretical optima.
	
	\item \textbf{Decoupled}: A prediction model is first trained targeting on the prediction accuracy, and then used to generate predictions of uncertain variables for the constrained optimization \eqref{eq:p}, corresponding to conventional predict-then-optimize framework.
	
	\item \textbf{End-to-End}: A prediction model  and constrained  optimization \eqref{eq:p} are integrated and trained in  an  end-to-end manner, corresponding to the proposed predict-to-optimize approach. 
\end{itemize}

For the prediction-based \textbf{Decoupled} and \textbf{End-to-End} methods, we evaluate both the prediction accuracy of uncertain variables and the operational performance of the IES. For the two methods, we first compute the optimal 
strategies based on the predictions, we then evaluate their performance using the realizations of uncertain variables.
Particularly,  deficient multi-energy supply  may occur   due to imperfect predictions. 
To address this issue, the deficient multi-energy supply is evaluated as penalty using the electricity price,  and surplus multi-energy supply is discarded.  
We therefore evaluate the \emph{ex-post}  operational performance of IES by the total cost:
\begin{equation}
	\begin{split}
		& L^{\rm total} \!\!=\! \sum_{ t \in T} \left[\lambda^{\rm b}_{t} P_t^{\rm g, buy} \!\!-\!\! \lambda^{\rm s}_{t} P_t^{\rm g, sell}  + \lambda^{\rm h}_{t} m^{\rm buy}_{t}  \right]\Delta t + \sum_{ t \in T} \lambda^{\rm b}_{t} \cdot D_t 
	\end{split}
\end{equation}
where $D_t$ denotes the deficient multi-energy supply, which can be evaluated according to \eqref{eq:energy_balance}.

We consider four case studies (\textbf{Case 1-4})
with increasing energy consumption levels of buildings and DCs. 
 Particularly, their peak multi-energy demand are scaled by factors of 0.1, 0.5, 1.0, and 1.5, respectively.
 Both the prediction accuracy of uncertain variables and operational performance of IES are evaluated and  reported   in TABLE~\ref{tab:performance}.
The prediction accuracy is evaluated by the widely-used performance metrics: MAPE, RMSE, and $R^2$ computed on normalized data, together with a regression loss calculated by MAE based on denormalized outputs.
From TABLE~\ref{tab:performance}, we observe that  the proposed \textbf{End-to-End} approach achieves similar prediction accuracy to the \textbf{Decoupled} method as indicated by all the performance metrics. 
However, the operational performance of the IES indicated by the total cost is  improved by approximately 7-9\%  by  the proposed  \textbf{End-to-End} approach over the \textbf{Decoupled} method. 
These improvements are noteworthy as we do not make any modifications on prediction models and  achieved only through the proposed end-to-end training scheme. 
Further, by comparing \textbf{End-to-End} approach with the \textbf{Optimal}, we find that the operational performance gap is around 19.5-33.4\% for  Case 2-4.  For Case 1, the performance gap is not available due to the zero total cost with \textbf{Optimal}, which is caused by the relatively low multi-energy demands that are fully satisfied  by on-site renewable energy and do not incur any grid purchase.


\begin{figure}[htbp]
	\centering
	\includegraphics[width=0.9\linewidth]{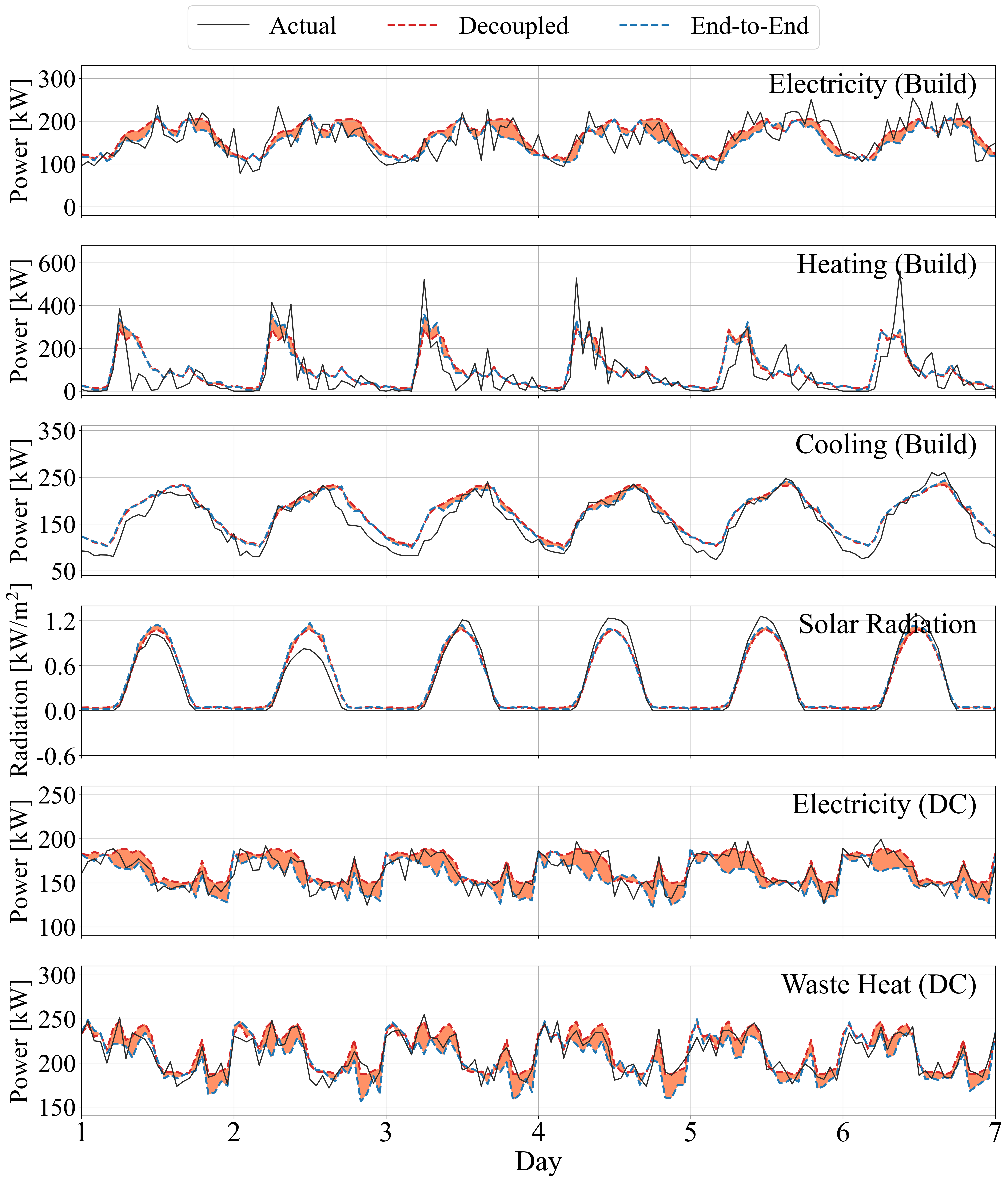}
	\captionsetup{belowskip=-14pt}
	\caption{Predictions of uncertain variables with the Decoupled and End-to-End methods for Case 3}
	\label{fig:prediction}
\end{figure}

The operational performance improvements of \textbf{End-to-End} over  \textbf{Decoupled} originate from the trained prediction models yield by the two methods. This can be perceived from Fig.~\ref{fig:prediction}, which shows 
obvious differences of the predictions yielded by the trained prediction models of the two methods.
Specifically, while the differences for heating and cooling demands, solar radiation are minor, there exists obvious deviations between the electricity demands of buildings and DCs, as well as the waste heat of DCs. 
These prediction discrepancies  actually lead to the operational performance gaps. 
However, by comparing the predictions with  actual values, we find that it is hard to discern which model provides more accurate predictions.  
We therefore infer that the end-to-end learning-based approach does not necessarily improve prediction accuracy. Instead,  they   guide the training of  prediction models to yield improved operational performance.



\begin{figure}[htbp]
	\centering
	\includegraphics[width=0.9\linewidth]{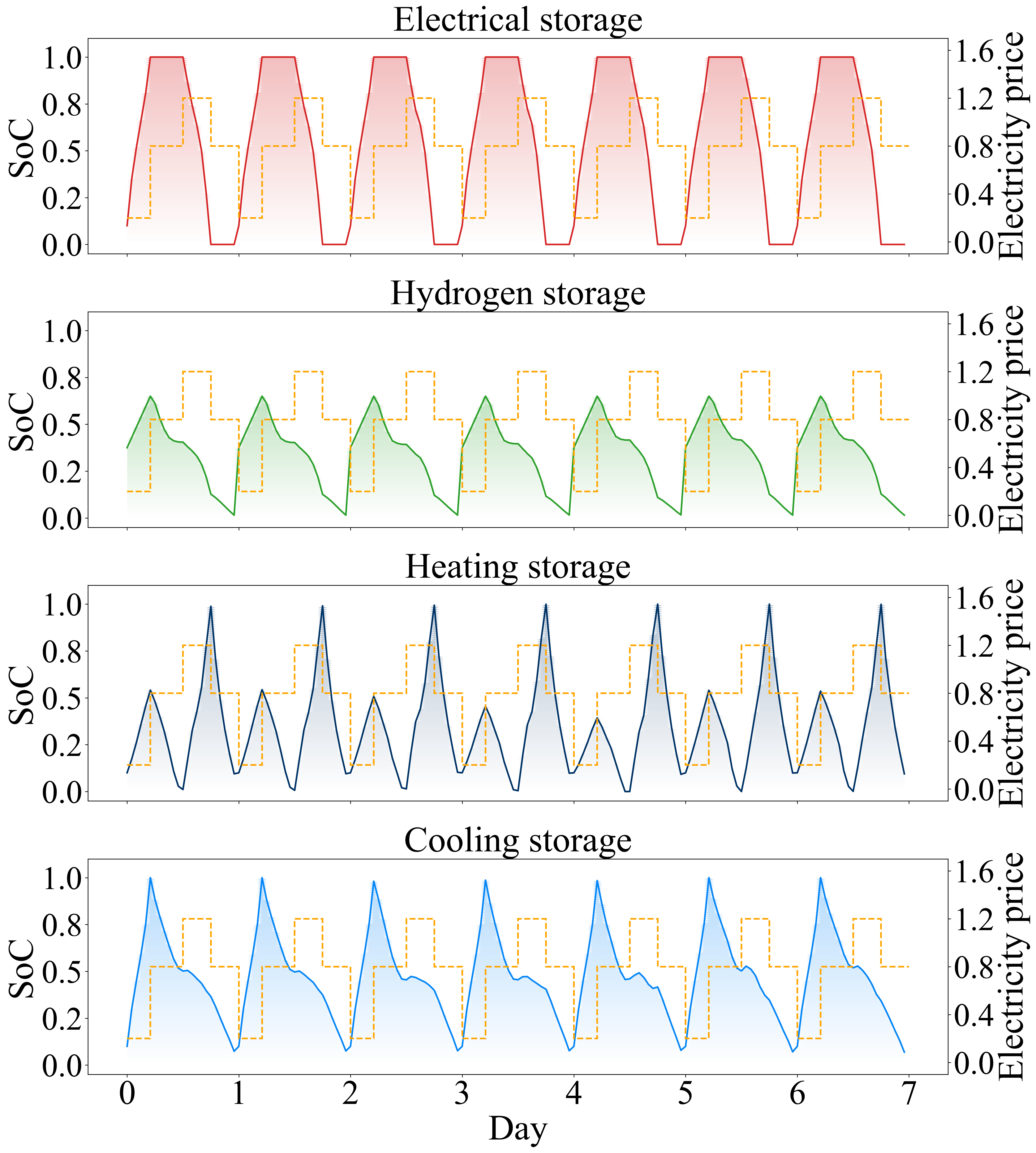}
	\captionsetup{belowskip=-10pt}
	\caption{Evolution of  energy storage devices with  End-to-End method in response to dynamic electricity prices for Case 3}
	\label{fig:soc}
\end{figure}


The effectiveness of operating strategies obtained from the proposed \textbf{End-to-End} method are further examined. Specifically, we investigate the evolution of  state-of-charge (SOC) of the multi-energy storage devices in response to the dynamic electricity price in Case 3.  As shown in Fig.\ref{fig:soc}, we observe all storage devices effectively respond to the dynamic electricity price. Specifically, they generally charge at low electricity price and  discharge when  high price arrives. This is reasonable and expected considering the energy cost reduction objective.

\vspace{-10pt}

\subsection{Economic benefits of waste heat recovery of DCs}
This section evaluates the economic benefits  of waste heat recovery of DCs within the IES. 
We consider varying workload levels of DCs and investigate how the waste heat recovery affects the total  cost of the system. 
Specifically, we vary the DC workload   from $20\%$ to $100\%$ with a fixed stepsize of 20\%, and assume that the electricity consumption and cooling demand of DCs are approximately proportional to the workload level. 
We compute the optimal operational strategy of the IES under different workload levels with and without  waste heat recovery based on the proposed end-to-end learning-based approach.  
We then evaluate the total cost (i.e., cost + penalty)  of the IES based on the obtained operating strategies. 

From  TABLE \ref{tab:economics}, we observe that when DC workload level exceeds 60\%, the total cost  is reduced by approximately 10\%.  In contrast, when the DC workload is below 40\%, the cost reduction is relatively limited. This is because when the DCs operate at high workload level, substantial waste heat is generated and can be reused, and vice verse when the DC workload level is low. 
We therefore conclude that by coordinating buildings and  DCs in IES and 
 effectively utilizing the generated waste heat of workload-intensive DCs, can substantially improve the overall energy efficiency.



\begin{table}[t]
	\centering
	\caption{Economic benefits of  waste heat recovery (WHR) of DCs on the IES.}
	\setlength{\tabcolsep}{3pt}
	\label{tab:economics}
	\begin{tabular}{c r r r}
		\toprule
		\makecell[c]{DC workload\\~[\%]} & \makecell[c]{Total cost w/o  WHR\\~[\$]} &  \makecell[c]{Total cost with  WHR\\~[\$]} & \makecell[c]{Reduction\\~[\%]} \\
		\midrule
 20  & \makecell[c]{2647.11} & \makecell[c]{2628.01} & \makecell[c]{0.7} \\
 40  & \makecell[c]{3435.88} & \makecell[c]{3323.71} & \makecell[c]{3.3} \\
 60  & \makecell[c]{4529.40} & \makecell[c]{4109.27} & \makecell[c]{9.3} \\
 80  & \makecell[c]{5534.78} & \makecell[c]{4934.78} & \makecell[c]{10.8}  \\
 100 & \makecell[c]{6594.32} & \makecell[c]{5732.46} & \makecell[c]{13.1} \\
		\bottomrule
	\end{tabular}
	\vspace{-12pt}

\end{table}

\vspace{-8pt}

\section{Conclusion}
\label{sec:conclusion}
This paper investigated an integrated energy system (IES) for coordinated multi-energy supply (i.e., electricity, heating and cooling) of buildings and data centers (DCs) with the objective to their synergistic  potential to improve energy efficiency. 
To address the effects of multi-energy supply and demand prediction errors on  the operational optimization of the IES, we proposed an  end-to-end learning-based  approach, which jointly integrates the training of prediction models for uncertain variables and constrained optimization in an end-to-end pipeline, as opposed to conventional predict-then-optimize paradigm.
Case studies based on real-world datasets show that the proposed method  improved the operational performance of IES by about 7-9\% over the predict-then-optimize methods. In addition,  coordinating the multi-energy supply of buildings and DCs show substantial economic benefits, and in particular, by recovering and reusing the waste heat of DCs to  meeting the heating and cooling demands of buildings and DCs can  significantly reduce the total energy cost of the IES.

\vspace{-10pt}

\bibliographystyle{ieeetr}
\bibliography{reference}

\end{document}